\documentclass[12pt,titlepage]{article}
\usepackage{graphicx}
\usepackage{amssymb}

\font\medio=cmr10 scaled \magstep2
\outer\def\beginsection#1\par{\medbreak\bigskip
      \message{#1}\leftline{\bf#1}\nobreak\medskip
\vskip-\parskip
      \noindent}

\def\laq{\raise 0.4ex\hbox{$<$}\kern -0.8em\lower 0.62
ex\hbox{$\sim$}}
\def\gaq{\raise 0.4ex\hbox{$>$}\kern -0.7em\lower 0.62
ex\hbox{$\sim$}}
\def\beq{\begin{equation}}
\def\eeq{\end{equation}}
\def\bea{\begin{eqnarray}}
\def\eea{\end{eqnarray}}
\def\bean{\begin{eqnarray*}}
\def\eean{\end{eqnarray*}}

\begin{document}
\bibliographystyle {unsrt}
\titlepage
\begin{flushright}

\vspace{15mm}
CERN-PH-TH/2012-350 \\

\end{flushright}
\vspace{15mm}
\begin{center}
{\Large \bf  Quantum hair and the string-black hole correspondence}\\

\vspace{15mm}

\vspace{6mm}
{ \bf  G. Veneziano$^{(1)}$}
\vspace{6mm}

{\sl Coll\`ege de France, 11 place Marcelin  Berthelot \\
 75005 Paris, France$^{(2)}$}\\ 
\vspace{3mm}
{\sl Center for Cosmology and Particle Physics \\
Department of Physics, New York University \\
4 Washington Place, New York, NY 10003,  USA} \\
\vspace{3mm}
{\sl Theory Division, CERN, CH-1211 Geneva 23, Switzerland}

\end{center}

\vskip 2cm
\centerline{\medio  Abstract}
\vskip 5mm
\noindent
We consider a thought experiment in which an energetic massless  string probes a ``stringhole" (a heavy string lying on the correspondence curve between strings and black holes) at large enough impact parameter for the regime to be under theoretical control. The corresponding, explicitly unitary, $S$-matrix turns out to be perturbatively sensitive to the microstate of the stringhole: in particular, at leading order in $l_s/b$,
it depends on a projection of the stringhole's Lorentz-contracted quadrupole moment. The string-black hole correspondence is therefore violated if one assumes quantum hair  to be exponentially suppressed as a function of black-hole entropy. Implications for the information paradox are briefly discussed.
\vspace{5mm}

\vfill
\begin{flushleft}
(1) e-mail: gabriele.veneziano@cern.ch

(2) Permanent address.

\end{flushleft}

\newpage
\section{Introduction}

 The issue of a possible loss of quantum coherence in processes in which a black hole is produced and then evaporates has been the subject of much debate since Hawking's claim \cite{Haw} that black holes should emit an exactly thermal spectrum of light quanta (see e.g. \cite{Mathur} for a recent review).
 Progress from string theory  on the microscopic understanding of black-hole entropy \cite{entropy}  and on 
the AdS-CFT correspondence \cite{AdS}, has lent strong support  to the belief that no loss of information/quantum-coherence should occur.
However, even in the AdS/CFT case,  understanding  how unitarity on the CFT side teaches about  information recovery on the gravity side remains unclear (see \cite{Malda}, \cite{BR}).

Another ``ab initio" approach to the same problematics  consists of the study of trans-planckian-energy collisions of massless strings as a function of center of mass  energy (or of the associated gravitational radius $R$), of impact parameter $b$, and of the string-length scale $l_s$, with the  relative ratios of these scales defining different regimes for the process  \cite{ACV1}.

 In this framework it has been possible to recover, within a unitarity-preserving $S$-matrix, both General Relativity expectations and string-size related modifications of it \cite{ACV2}, albeit  in regimes in which no-black hole formation is expected  according to closed-trapped surface criteria\cite{CTS}. Dealing with the complementary regime (corresponding to  $R \gg b, l_s$)   has met with more limited success, although some progress has been made in understanding how the threshold of black-hole production can be approached from below \cite{GV04}. An approximation to deal with the full-collapse regime, proposed a few years ago in \cite{ACV07}, appears to predict correctly the existence and rough values of some critical ratios for the onset of collapse, but, unfortunately, has failed so far to provide a unitary description of the process beyond such critical points \cite{CC}.

Given the above  difficulties, the attention has been shifted to a supposedly easier problem \cite{DDRV}, that of the scattering of a closed light string off a stack of $N$ $D-p$-branes at small string coupling and large $N$. Here the equivalent of the black-hole formation regime is  the one in which the closed string is absorbed by the brane system and its energy is dissipated in open string excitations of the stack itself. In spite of some progress \cite{DDRV} \cite{MWB}, understanding how information about the initial state gets encoded in the final one is still far from settled.

One problem is that information, if it's to be eventually recovered, has to start coming out, at the latest, by the so-called Page time \cite{Page}, corresponding roughly  to the time by which the evaporating black hole has lost half of its entropy $S$.  In order for this to be possible, the rate of information retrieval  cannot be too small, e.g. cannot be of order $\exp(-S)$, at least  not after the Page time. Information retrieval  should  instead  be easy if ``quantum hair" is inversely proportional to $S$,
as recently proposed in a toy model identifying black holes with a self-sustained critical Bose-Einstein condensate of $N \sim S$ gravitons \cite{DG}. Similar claims have been made in \cite{Ramy} on the basis of general uncertainty-principle considerations applied to the geometry itself. Indeed, once an effective classical geometry with an information-free horizon is assumed (even an effective one that corrects the classical horizon), continuous information loss looks  inevitable\cite{Mathur}.

In this paper we will address this kind of questions using the correspondence between strings and black holes \cite{Corr1, HP, DV} that occurs when the mass of the former is tuned to the value $M_{SH} = M_s g_s^{-2}$, giving a Schwarzschild radius $R = O(l_s)$. By going to small enough string coupling we can make the entropy of such ``string-holes" (SH) arbitrarily large:
\beq
\label{SSH}
S_{SH} =  \left( \frac{l_s}{l_P}\right)^{D-2} =  \left( \frac{M_P}{M_s}\right)^{D-2} = g_s^{-2} \gg 1 \, .
\eeq

It is particularly appealing that, for SHs, the question of the size of quantum hair becomes one about whether it is perturbative or not in the string coupling constant. In our case, the role of the parameter $N$ of \cite{DG} is played by the string coupling which, for a given string mass,  is tuned to a critical value. Unfortunately, and unlike in the simple model of \cite{DG}, we are presently unable   to perform a reliable calculation when $g_s$ and/or $M$ are parametrically larger than their critical values.

Furthermore, in order be able to claim that strings of mass $M_{SH} = M_s g_s^{-2}$ can also be seen as black holes, we have to impose that they are compact enough not to exceed  in size their own Schwarzschild radius $R = O(l_s)$, and to check that this restriction does not invalidate the entropy estimate (\ref{SSH}).
This question was addressed in \cite{DV} (see also \cite{HP}), where it was argued that the entropy of string states of mass  $M \le M_{SH} = g_s^{-2} M_s$ is shared among states of different size $r$ according to:
\beq
\label{S(M,R)}
S(M, r) \sim \frac{M}{M_s} \left( 1- c_1 \frac{l_s^2}{r^2} \right)  \left( 1- c_2 \frac{r^2}{(\alpha' M)^2} \right)\left(1 +c_3 \left(\frac{R}{r}\right)^{D-3} \right)\, ,
\eeq
where $c_i$ are positive constants of $O(1)$. For $M \ll M_{SH}$ the last terms is negligible and the first two factors give a maximal entropy for
$r \sim l_s \sqrt{\frac{M}{M_s}}$, the random-walk value. However, there is still an entropy $O(M/M_s)$ in ``compact" strings and, furthermore, as one approaches
$M= M_{SH}$, the third term in (\ref{S(M,R)}) helps favoring such strings.

Another way of reaching a similar result consists in counting string states at level $N = \alpha' M^2$ produced by oscillators of index larger than $K$.
A simple argument, based on evaluating the corresponding partition function, shows that the entropy of such states is still $O(\sqrt{N})$ if $K \sim \sqrt{N}$.
They will generally correspond to occupation numbers $\le O(1)$ for $O(\sqrt{N})$ oscillators (providing the right value for their mass) and will have a size of order:
\beq
r^2 \sim l_s ^2 \sum_{n > \sqrt{N}} \frac1n \langle  a_n^{\dagger} a_n  \rangle \sim{l_s^2}\, .
\eeq
This is  the kind of states we shall focus our attention on. We recall that, not only entropy, but, qualitatively, many other properties of strings and black holes (decay rates, evaporation time etc.) match on the correspondence line  \cite{HP,DV}.

The idea, therefore,  is  to  consider a thought experiment in which a massless string probes a stringhole target, a process somewhere in between those discussed in \cite{ACV1} (where both projectile and target are massless) and \cite{DDRV} (where the target is infinitely heavy). Studying such a process at sufficiently large impact parameters for the approximations to be under control turns out to be sufficient to reveal whether the quantum hair of such SHs is perturbative or not in 
$1/S \sim g_s^{2} $. This appears to be the string-theory counterpart to checking (albeit only at a specific point) whether  quantum hair is perturbative in $1/N$ in the approach of \cite{DG}.

\section{ A thought experiment revealing quantum hair}

 We work in flat  $10$-dimensional spacetime with  $(10- D)$ dimensions compactified at the string-length scale so that the effective large-distance physics lives in $D$ spacetime dimensions. We are also assuming to be working at very small string coupling $g_s$ so that, as already indicated in (\ref{SSH}),  there is a large hierarchy between the string and Planck mass scales.

Consider now a process in which a  massless ``probe" string collides with a well-defined heavy (and for the moment generic) ``target" string of mass $M \gg M_P \gg M_s$. Let us also take a high-energy limit in which the energy $E$ of the probe string  in the rest frame of the heavy one is much larger than $M_s$ and yet much smaller than $M$, 
\beq
\label{Ebounds}
M_s M \ll s - M^2 =  -2 p\cdot P = 2 E M \ll M^2\, ,
\eeq
so that the light string does indeed act (almost) as a probe and yet we can apply a high-energy limit in which graviton exchange dominates.

Following the logic of \cite{ACV1} (see also \cite{Iengo}, \cite{DDRV}) we can argue that, at large-enough impact parameter $b$, the elastic  scattering amplitude is  given by the semiclassical eikonal formula:
\beq
\label{classps}
\label{leadeik}
S(E, M, b) \sim \exp(i \frac{{\cal A}_{cl}}{ \hbar}) = \exp\left(i \frac{4G E M}{\hbar} c_D b^{4-D}\right) \equiv e^{2i\delta(E,M,b)}~;~ c_D = \Omega_{D-4}^{-1} \equiv \frac{\Gamma(\frac{D-4}{2})}{2 \pi^{\frac{D-4}{2}}}.
\eeq
As a consistency check, we note that, when one goes back from $b$ to $q$-space (or deflection angle $\theta$), one recovers, at the saddle point of the $b$-integral, the classical  Einstein relation (generalized to arbitrary $D$) between deflection angle, mass, and impact parameter:
\beq
\label{Einstein}
\theta = \frac{8 \pi G M}{\Omega_{D-2} b^{D-3}} \sim \left(\frac{R}{b}\right)^{D-3} \ll 1  ~~;~~ (G M)^{\frac{1}{D-3}} \sim R \ll b \, ,
\eeq
where $R$ is the Schwarzschild radius of the heavy string. Obviously, the above formula satisfies the ``no-hair" theorem, in the sense that it is sensitive to the mass  of the heavy string state but not to its microscopic quantum numbers.

Diagrammatically, the result (\ref{leadeik})  comes from exponentiating the exchange of a single graviton between the light and the heavy string. Both (\ref{classps}) (and (\ref{Einstein})) are indeed only valid at sufficiently large impact parameter (small  deflection angle) and suffer from corrections of higher order in $R/b$ ($\theta$). These will reconstruct, for instance, the  deflection formula in the full Schwarzschild (or Kerr if we consider a target with spin) metric. As shown long ago by Duff \cite{Duff}, they correspond, diagrammatically, to exponentiating connected graviton-tree (fan) diagrams in which a single vertex (the trunk of the tree) is attached to the probe string while all the branches terminate on the heavy one, giving the appropriate powers of $R$ and $b$. These classical correction still satisfy the no-hair condition as well as elastic unitarity. On the other hand if instead``hairy" corrections to $\delta(E,M,b)$ are exponentially suppressed for large black holes, we would expect them to show up in the form:
\beq
\label{ExpSuppr}
\delta(E,M,b) \rightarrow \delta(E,M,b)(1+{\rm classical ~corrections} + e^{-c S} \hat{Q} )  \, ,
\eeq
where $c$ is some constant and $ \hat{Q} $ represents, schematically, a quantum-hair operator taking different expectation values depending on the  black hole microstate. We will check below whether an ansatz like (\ref{ExpSuppr}) is satisfied for the particular stringhole states introduced in the previous Section.

To address this question recall that, as discussed in \cite{ACV1} and \cite{DDRV} in two different contexts, there are also ``string corrections" to the leading eikonal form. These are related to the fact that strings are extended objects and therefore suffer tidal forces when moving in a non trivial geometry\cite{Giddings}\footnote{Although all calculations are performed in  flat spacetime the effects of an effective non-trivial geometry emerge from the calculation.}. Fortunately, at least at small scattering angle, such corrections are fully under control and lead to a unitary $S$-matrix.
Unitarity is now satisfied in a less trivial way: different channels couple, elastic unitarity is violated, but one still obtains a fully unitary $S$-matrix in the Hilbert space of two arbitrary string states. The  question is whether this non-trivial S-matrix contains information about the actual state of the heavy string, and at which level.

Building on the work of  \cite{ACV1} and \cite{DDRV}  we can be confident that the tidal excitation of both the light and the heavy string are captured, at leading order in $\theta$, by the replacement:
\beq
\label{quantumps}
\delta(E, M, b)  \rightarrow   \hat{\delta}(E, M, b) =    \langle \delta (b + \hat{X}_H - \hat{X}_L) \rangle  = 2 G E M \hbar^{-1} c_D \langle (b + \hat{X}_H - \hat{X}_L)^{4-D} \rangle \, .
\eeq
Here $\hat{X}_H$ and $\hat{X}_L$ represent the heavy and light string position operators, stripped of their zero modes (which give $b$), evaluated at $\tau =0$, and averaged over $\sigma$. These operations, together with a normal-ordering prescription, are indicated in (\ref{quantumps}) by the brackets, i.e.
\beq
 \langle (b + \hat{X}_H - \hat{X}_L)^{4-D} \rangle  \equiv \frac{1}{4 \pi^2} \int_0^{2 \pi} d \sigma_L   \int_0^{2 \pi} d\sigma_H : \left(b + \hat{X}_H(\sigma_H, 0) - \hat{X}_L(\sigma_L, 0)\right)^{4-D} : ~ .
\eeq
 In  words, the classical phase shift is  replaced by the average of a {\it quantum} phase shift in which the impact parameter is affected by a quantum uncertainty encoded in the string position operators.
 
For what concerns the excitation of the light string,  further justification of the above formula comes from the study of string-brane collision discussed in \cite{DDRV}, specialized to the case of a stack of $0$-branes. 
For the excitation of the heavy string we can instead appeal to the quantization of the heavy string in the shock-wave metric produced by the light one  \cite{GGM}.

Following  \cite{ACV1}, we now expand (\ref{quantumps}) to quadratic order in the $\hat{X}$ (the linear order clearly averages out to zero) to get the leading correction in an  expansion in $(l_s/b)^2$:
\beq
\label{quantumps2nd}
2( \hat{\delta} - \delta)  =  \frac{2 \pi G E M (D-2)}{\hbar \Omega_{D-2} b^{D-2}} \langle Q_H^{ij} + Q_L^{ij}  \rangle \hat{b}_i  \hat{b}_j \, .
\eeq
Here $Q_H^{ij}$ is the $(D-2)$-dimensional (i.e. Lorentz-contracted in the direction of the incoming momentum) quadrupole operator for the heavy string\footnote{I am  grateful to T.  Damour for this interesting remark.}.
\beq
\label{Q}
Q_H^{ij} =  \hat{X}_H^i  \hat{X}_H^j  - \frac{ \delta_{ij}}{D-2} \sum_{i=1}^{D-2}  \hat{X}_H^i  \hat{X}_H^i \, ,
\eeq
and  is projected along the unit vector  $\hat{b}$ in the direction of the impact parameter. This projection can also be written in the form:
\beq
\label{Pi}
Q_H^{ij}  \hat{b}_i  \hat{b}_j  =   \hat{X}_H^i  \hat{X}_H^j  \left( \hat{b}_i  \hat{b}_j  -\frac{ \delta_{ij}}{D-2} \right) \equiv  \Pi_{ij}  \hat{X}_H^i  \hat{X}_H^j \, .
\eeq
As indicated in (\ref{quantumps2nd}), we get a similar term for the probe string. At this order in $l_s/b$ the $S$-matrix thus factorizes in the form:
\beq
\label{quantumS}
S(E, M, b)  =  \exp (2i\delta)~ \Sigma_L~ \Sigma_H~;~ \Sigma_{L,H} = \exp \left(i (D-2) \Delta~ \tilde{Q}_{L,H}^{ij} ~ \hat{b}_i  \hat{b}_j  \right)\, ,
\eeq
where we have defined the dimensionless quantities\footnote{The value of $\Delta$, when compared to unity, determines \cite{ACV1} whether the probe string gets excited or not by tidal forces. However, once more, these effects will {\it not} depend on the particular state of the target string.}:
\beq
\label{Delta}
\Delta =   \frac{2 \pi G E M  l_s^2}{\hbar \Omega_{D-2} b^{D-2}} ~~ \, ; \, ~~ \tilde{Q}^{ij} = l_s^{-2} Q^{ij}\, ,
\eeq
the latter being the quadrupole measured in string-length units.
Since the quadrupole operators are hermitian (see also below), each factor appearing in (\ref{quantumS}) corresponds to a unitary operator. The first two factors
are  independent of the particular state chosen for the heavy string. Let us therefore concentrate our attention on $\Sigma_H$ (dropping for simplicity the $H$ suffix).  The operator appearing at the exponent in $\Sigma$ can be easily written down:
\beq
\label{quantumps2ndosc}
\tilde{Q}^{ij} ~ \hat{b}_i  \hat{b}_j=  \Pi_{ij} ~  \sum_{n=1}^{\infty}  \frac1n\left( a^{ \dagger i}_n a_n^j + \tilde{a}^{ \dagger i}_n \tilde{a}_n^j + a_n^i  \tilde{a}^{ j}_n +  a^{ \dagger i}_n  \tilde{a}^{ \dagger j}_n  \right)\, .
\eeq
 Its diagonal matrix elements are  sensitive to the (projected, transverse) quadrupole of the heavy string, while the transitions to other states,  induced by terms with two creation or two annihilation operators,  correspond to a quadrupole-like excitation of the original string itself. This is hardly surprising in view of the intimate relation between tidal forces and quadrupole moments (see e.g. \cite{tidalQ}), and simply appears as a generalization of known facts to an ultra-relativistic situation involving strings (our quadrupole, in particular, is a purely geometrical object). 

In order to have an estimate of quantum hair we need to normal order the whole exponential operator  occurring in  $\Sigma$. Following again \cite{ACV1}, we find:
\bea
\label{NOSmatrix}
\Sigma_H &=&  \Sigma^{(univ)}~ \Sigma^{(hair)}~;~
\Sigma^{(univ)} = \Gamma (1+ i \Delta)^{D-3} ~  \Gamma (1-  i (D-3) \Delta)  \nonumber \\
  \Sigma^{(hair)} &=& : \exp  \left( \sum_{n=1}^{\infty}  (a^{ \dagger i}_n +  \tilde{a}_n^i)(a_n^j +  \tilde{a}^{ \dagger j}_n)\left[ C_n(\Delta)(\delta_{ij} -   \hat{b}_i  \hat{b}_j ) +  \tilde{C}_n(\Delta)  \hat{b}_i  \hat{b}_j \right] \right):   \nonumber \\
   C_n(\Delta) &=& - \frac{i \Delta}{n + i \Delta} ~;~ \tilde{C}_n(\Delta)  =  C_n(- (D-3) \Delta)\, .
\eea
At this point the explicit calculation of the $S$-matrix is simple, in particular between coherent states. $\Sigma^{(univ)}$, being a $c$-number,  does not depend on the internal quantum numbers of the heavy string and, together with similar factors coming from the light string,  provides absorption and further contributions to the phase shifts, but no hair. Instead, the operator  $\Sigma^{(hair)}$ generates matrix elements that feel the nature of the microstate in which the heavy string actually  is. Note that normal ordering has slightly upset the exact quadrupole structure appearing in (\ref{Q}), (\ref{Pi}) (which is however recovered for  $n \gg \Delta$).

Let us now specify further the process described in the previous section in order to make contact with black-hole physics. To this purpose we shall identify the heavy string with a ``stringhole" state described in Sec. 1. 
The reason for choosing that precise (within factors $O(1)$) value of $M$ is twofold. Choosing $M$ in the range $M_P \ll M \ll M_{SH}$ leads to reliable results, but the string, in this case, is below the correspondence curve, its size is larger than its Schwarzschild radius  and therefore is not a collapsed object \cite{Corr1}.
On the other hand, various approximations that can be justified for strings of mass up to $M_{SH}$ cease to be valid for strings with $M >> M_{SH}$, i.e. strings that would simulate ``large" black holes in string-length units. 

Let us first evaluate the quantity $\Delta$ in (\ref{Delta}) for the SH case. Up to numerical factors:
\beq
\Delta =  \frac{G E M  l_s^2}{\hbar ~b^{D-2}}  \rightarrow  \frac{E  l_s}{\hbar} \left(\frac{ l_s}{b}\right)^{D-2} \sim \frac{E}{M_s} \theta^{\frac{D-2}{D-3}} \, .
\eeq
Given our bounds (\ref{Ebounds}) on $E$ we find:
\beq
 \theta^{\frac{D-2}{D-3}} \ll \Delta \ll g_s^{-2}  \theta^{\frac{D-2}{D-3}} \, .
\eeq
Obviously, even keeping $\theta \ll 1$, but finite and $g_s$-independent, we can make $\Delta \gg 1$ (yet $\ll g_s^{-2}$) for sufficiently small $g_s$ and with $E$ in a parametrically large region.

  In order to estimate the size of quantum hair we note that the coefficients $C_n(\Delta)$  appearing in (\ref{NOSmatrix}) become $O(1)$ at  $n < \Delta$ or of order
$\Delta/n$ at $n > \Delta$.  As already discussed, typical SHs will have most of the non vanishing occupation numbers  of $O(1)$  in oscillators with  $n \sim \sqrt{N} \sim g_s^{-2}$.
In that case $C_n \sim \Delta/n$ and  eq. (\ref{NOSmatrix}) simplifies further:
\beq
\label{simpler}
  \Sigma^{(hair)} = : \exp  \left( - i (D-2) \Delta \sum_{n=1}^{\infty}\frac{1}{n}  (a^{ \dagger i}_n +  \tilde{a}_n^i)(a_n^j +  \tilde{a}^{ \dagger j}_n) \Pi_{ij} \right):  ~.
\eeq 

The basic observation is that the operator appearing in the exponent of (\ref{simpler}) is completely unrelated to the one giving the mass of the SH\footnote{It is also clearly non-degenerate with the spin of the SH.} and therefore will distinguish degenerate microstates. It contains  positive definite diagonal terms that correspond to the transverse, projected quadrupole of the SH. The non-positive definite terms, corresponding to inelastic transitions, are also microstate-dependent through a similar quadrupole operator. There is also a state-dependent absorption from the real part of $C_n$. This is suppressed by an extra factor $\Delta/n \sim g_s^2 \Delta$ and is not  controlled  by the quadrupole.

The dominant terms sum up to something $O(1)$ but can still take different values of that same order within the whole SH ensemble. Hence, an experiment measuring the phase of the $S$-matrix should be able to reduce our ignorance on the state of the SH by a factor $O(2)$\footnote{If instead we wish to distinguish SH states with differences $O(1)$ in the occupation numbers  the sensitivity of (\ref{simpler}) will have a suppression factor\ of $O(\Delta/n) \sim  g_s^2  \Delta$.}. Interpreting this as a reduction on the total number of states $e^S$, it will correspond to a decrease of $O(1)$  in entropy, meaning that the whole information can  be recovered after $O(S)$ experiments. The minimal duration of each experiment being $O(l_s)$, namely the light-crossing time for a SH,  the total time needed to recover the information will be of order of the Page/evaporation time $S R \sim g_s^{-2} l_s$. 

 In our  approximation quantum hair also appears to be suppressed with respect to the no-hair terms by a power of the scattering angle.  While this is still sufficient for our qualitative discussion, we think that our results should  qualitatively extend to scattering angles of O(1). Checking this is not easy since, precisely for a SH target, string-size and classical corrections kick in simultaneously as we increase the scattering angle (although the different $\theta$-dependence should help separate the two kinds of corrections).
If so, the quantum hair revealed by our scattering process (with higher multipoles appearing besides $Q_{ij}$)  will indeed approach $O(1)$ for a probe-energy of order of the Hawking/Hagedorn temperature of the SH. Such impact parameters and energies are precisely those typical of Hawking's radiation. 
  Actually, as well known in particle physics (see e.g. \cite{SW}), a decay amplitude is usually to be corrected by a ``final-state interaction" which basically amounts to  multiplying the naive decay amplitude by a factor $S^{1/2} (E,b) \sim \exp(i \delta(E,b) )$, where the typical values of $E$ and $b$ will be
 $ M_s$ and $l_s$ respectively.
  In other words, such a quantum hair may  directly leave its imprint in the decay of a SH.
  Admittedly, all these are hand-waving arguments that should be analyzed more carefully.

   In any case, it appears  that the quantum-hair amplitude is {\it not} suppressed, relative to no-hair terms, by $\exp(-S_{SH}) \sim \exp(-g_s^{-2})$ but rather, at most,  by a small inverse power of $S \sim g^{-2}$, i.e. is a perturbative effect in the string coupling. Note, however, that a generic individual element of the $S$-matrix is always suppressed by an $\exp{(-\Delta})$ ``non-perturbative" factor, which gets compensated by the exponentially large number of final states contributing to inclusive-enough cross sections.
 Indeed, given that $\Sigma_H$ is unitary, it is easy to lose its sensitivity to quantum hair if traces over the heavy initial and/or final SH states are taken. Summing individual transition probabilities  over  final SH states corresponds to considering an inclusive cross section, while  tracing/averaging over the initial SH state corresponds to an initial mixed state. In both cases it is quite clear that unitarity of $\Sigma_H$ washes out  all the leading-order SH hair discussed so far\footnote{At order $l_s^4 b^{-4}$ the eikonal operator will give terms proportional to $X_L^2 X_H^2$ that destroy factorization.}. Only subleading terms and/or appropriate interference experiments will be able to leave  information about the state of the SH on the probe-string. Whether this is in principle sufficient to retrieve enough information on the SH is not completely obvious.

In conclusion, the results we have presented point in the direction of some perturbative quantum hair being revealed in our thought experiment, very likely something of order $1/S \sim g_s^2$ for a probe of $E \sim M_s$ during a collision (horizon-crossing) time $O(l_s)$.
At least naively, this would allow to retrieve the full information about the microstate of the string hole  within its evaporation time  of order
$g_s^{-2} l_s$. This result can be  related to the fact that, for a stringhole, the concept of an infomation-free horizon does not make sense (the horizon being as large as the string itself) and, in this sense, it is similar to what is believed to occur for the so-called fuzzballs states of string theory (see \cite{fuzzballs} and references therein). It would be interesting to see whether thought experiments of the kind discussed here using fuzzballs could reveal a similar amount of quantum hair. Of course the issue of whether or not spacetime around the horizon can be considered to be empty is also very relevant in the recent firewalls debate \cite{firewalls}.

Can we reconcile our finding with an exponentially small amount of quantum hair for large black holes (i.e. for  black holes much heavier that $M_{SH}$, for which our simple analysis fails to provide a reliable answer)?   Clearly an expression like that of  (\ref{ExpSuppr}) is in contradiction with our findings but one could instead imagine an ansatz like:
\beq
\label{DiffSuppr}
\delta \rightarrow \delta(E,M,b)(1+{\rm classical ~corrections} + e^{-c \frac{S}{S_{SH}}} \hat{Q} )  \, ,
\eeq
which would only give an exponential suppression for black holes that are much heavier that those on the correspondence line.
Indeed, a single string may fail to represent black holes above the correspondence curve (seen in that case as a critical line separating two phases), in which case $\Sigma^{(hair)}$ could change quite abruptly above the phase transition.

Another  possible objection to drawing strong conclusions from our results lies in the possibility\footnote{This possible loophole was suggested by M. Porrati.} that SHs do {\it  not} represent typical black holes but only a tiny fraction of them. In that case, their long hair will make them atypical ``hippie-like" black holes within  a vast majority of ``bald" ones. 

\section*{Acknowledgements}
This investigation was prompted by a stimulating seminar by Gia Dvali and by subsequent discussions with him and Ramy Brustein. I have also benefitted from interesting discussions and/or correspondence with Daniele Amati,  Thibault Damour, Sergei Dubovsky, Gregory Gabadadze, Cesar Gomez, Hovhannes Grigoryan, Matthew Kleban, Samir Mathur, Yaron Oz, Massimo Porrati, Eliezer Rabinovici, Rodolfo Russo and Adam Schwimmer. I also wish to acknowledge the support of an NYU Global Distinguished Professorship.

\end{document}